\documentclass[adp,fleqn]{w-art}
\usepackage{times}
\usepackage{w-thm}
\usepackage[]{graphicx}
\chardef\bslash=`\\ 

\begin{document}
\DOIsuffix{theDOIsuffix}
\Volume{12}
\Issue{1}
\Copyrightissue{01}
\Month{01}
\Year{2003}
\pagespan{1}{}
\Receiveddate{xx Month 2005}



\title[Dynamic conductivity of manganites]{Dynamic conductivity of semiconducting manganites approaching the metal-insulator transition}


\author[P.\ Lunkenheimer]{P. Lunkenheimer\footnote{Corresponding
     author{\quad}E-mail: {\sf Peter.Lunkenheimer@Physik.Uni-Augsburg.de}}}
\address{Experimental Physics V, Center for Electronic Correlations and Magnetism,
University of Augsburg, 86135 Augsburg, Germany}
\author[F.\ Mayr]{F. Mayr}

\author[A.\ Loidl]{A. Loidl}

\begin{abstract}
  We report the frequency-dependent conductivity of the manganite
  system La$_{1-x}$Sr$_{x}$MnO$_{3}$ ($x\leq0.2$) when approaching the metal-insulator
  transition from the insulating side. Results from low-frequency dielectric measurements
  are combined with spectra in the infrared region. For low doping
  levels the behavior is dominated by hopping transport of
  localized charge carriers at low frequencies and by phononic and electronic
  excitations in the infrared region. For the higher Sr contents
  the approach of the metallic state is accompanied by the successive
  suppression of the hopping contribution at low frequencies and
  by the development of polaronic excitations in the infrared
  region, which finally become superimposed by a strong Drude
  contribution in the fully metallic state.

  \end{abstract}
\maketitle





\section{Introduction}
\label{sect1} In recent years, the perovskite-related manganites
have attracted tremendous interest, mainly triggered by the
discovery of the colossal magnetoresistance (CMR) \cite{cmr}. But
these systems, being governed by many competing interactions
(superexchange, double exchange, charge ordering, and Jahn-Teller
effect) leading to very rich phase diagrams, are also renowned for
providing prototypical examples of temperature, doping, and
magnetic-field induced metal-to-insulator (MI) transitions. The
theoretical \cite{furukawa,millis1,horsch} and experimental
\cite{oki1,kaplan,jung1,jung2,quijada,machida,seeger,oki2,take,pao1,mayr1,sichel,tobe,hart1}
investigation of the dynamical conductivity played a significant
role in the clarification of the charge transport processes and
the types of charge carriers in these manganites. However, still
our understanding of the many puzzling properties of these
materials is far from complete and the theoretical understanding
of the importance of the electron-phonon coupling
\cite{millis2,roder} has to be verified in optical experiments..
In the present work, we report on the evolution of the dynamic
conductivity of the classical CMR manganite
La$_{1-x}$Sr$_{x}$MnO$_{3}$ ($0 \leq x \leq 0.2$) when approaching
the MI transition from the insulating side. Both, dielectric
measurements in the Hz-GHz frequency range and optic measurements
in the far- to near-infrared range were performed. They provide
complementary information giving insight into the mechanisms that
are active when the metallic state, hallmarked by a well
pronounced Drude contribution, develops out of the insulating one,
characterized by strongly localized charge carriers. Our results,
together with those obtained on various other transition metal
oxides, allow for the development of a general scenario for the
evolution of the broadband dynamic conductivity when approaching
the metallic state from the insulating side.

\begin{vchfigure}[htb]
\includegraphics[angle=270,width=.6\textwidth,clip]{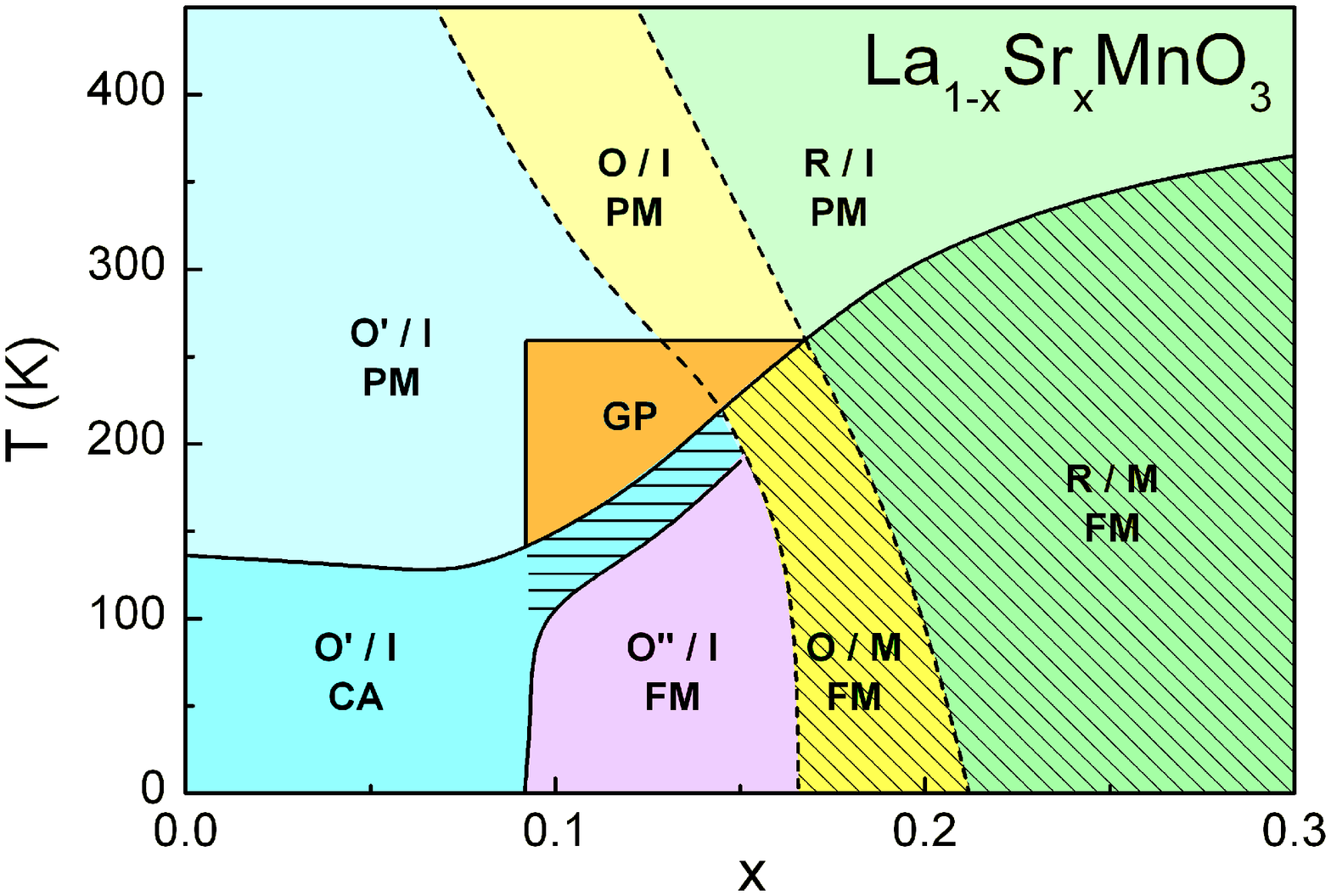}
\vchcaption{The ($x$, $T$) phase diagram of
La$_{1-x}$Sr$_{x}$MnO$_{3}$
\cite{ParaJMMM,ParaJPhys,Griffiths,hemb}. The structural (O,
O$^{\prime}$, O$^{\prime\prime}$, R), magnetic (PM, CA, FM), and
electronic (M, I) phases are indicated. The diagonally hatched
areas indicate the metallic states. In the horizontally hatched
area, electronic phase-separation may exist \cite{ParaJPhys}. "GP"
denotes the Griffiths phase, proposed in \cite{Griffiths}. }
\label{PhaseDiag}
\end{vchfigure}

LaMnO$_{3}$ is a charge-transfer insulator. In the octahedral
environment the 3$d$-levels of the  Mn$^{3+}$ ions are split into
a lower $t_{2g}$ triplet and an excited $e_{g}$ doublet. The
$e_{g}$ states are filled with one electron and hence are strongly
Jahn-Teller active. Indeed, below 800~K LaMnO$_{3}$ exhibits
orbital order with a half-filled lower and an empty excited
$e_{g}$ level. On substituting La$^{3+}$ by Sr$^{2+}$, holes are
introduced into the lower $e_{g}$ level, weakening the JT
interactions and driving the system into a metallic state
\cite{jonker}. Already in the early days it was speculated that
for the metal-to-insulator transition double exchange might play
an important role, driving the system from a paramagnetic
insulator into a ferromagnetic metal \cite{gennes}.

The rich phase diagram of the La$_{1-x}$Sr$_{x}$MnO$_{3}$ system
for low doping levels is given in Figure \ref{PhaseDiag}, based on
dc, magnetic, optic, and electron spin resonance measurements
\cite{ParaJMMM,ParaJPhys,Griffiths,hemb}. Depending on doping and
temperature, the system exhibits a rhombohedral (R) and three
different orthorhombic structures (O, O$^{\prime}$,
O$^{\prime\prime}$; see \cite{ParaJMMM,ParaJPhys,hemb} for
details). At low concentrations ($x<0.1$) there is a Jahn-Teller
(JT) distorted and insulating orthorhombic phase O$^{\prime}$,
which at low temperatures reveals canted antiferromagnetism (CA).
The ground state is an orbitally ordered and ferromagnetic
insulator (FM/O$^{\prime\prime}$/I) for $0.1 \leq x \leq 0.15$,
and a FM metal for $x>0.17$ in the non-JT distorted O and R phases
(diagonally hatched area). Finally, the triangular region, marked
"GP", denotes the Griffith phase, which was postulated in
\cite{Griffiths}, based on electron spin resonance and magnetic
measurements. The formation of this phase, which is characterized
by ferromagnetic clusters formed below a characteristic
temperature $T_{G}$ ($\approx 270$~K in the present case), was
ascribed to quenched disorder of the randomly distributed FM bonds
in this material \cite{Griffiths}. At the FM phase boundary, the
clusters achieve FM order and percolate.

\section{Experimental Details}
\label{sect2} Well characterized single crystals of
La$_{1-x}$Sr$_{x}$MnO$_{3}$, grown by the floating zone method,
were kindly provided by Profs. A.A. Mukhin and A.M. Balbashov
\cite{samples}. Measurements of the dc conductivity were performed
using a standard four-point technique. For the non-metallic
samples, the ac conductivity at frequencies below 1~GHz was
recorded using an autobalance bridge HP4284 covering frequencies
20~Hz~$\leq \nu \leq$ 1~MHz and an HP4291 impedance analyzer at
frequencies 1~MHz~$\leq \nu \leq$ 1.8~GHz \cite{ferro}. The
microwave conductivity was measured at 7.3 GHz utilizing a
microwave perturbation technique within a $^{4}$He-flow cryostat.
In the IR regime the temperature-dependent conductivity was
determined via reflectivity measurements using a Bruker IFS 113v
Fourier transform spectrometer. From the reflectivity $R$, the
conductivity $\sigma ^{\prime}$ was calculated via the
Kramers-Kronig transformation, requiring an extrapolation of the
data towards low and high frequencies. For $\nu \rightarrow 0$, a
constant extrapolation was used for the insulating states and a
Hagen-Rubens law for the metallic states. At high frequencies the
spectra were extrapolated by a power law $R\propto \nu^{-1.5}$ up
to $10^{6}$~cm$^{-1}$, leading to a smooth transition to the
experimental data, which was followed by a $\nu^{-4}$ power law at
higher frequencies \cite{mayrprom}. For some samples, additional
measurements at 7.3~GHz and around 100~GHz were performed using a
microwave perturbation technique and a quasi-optic Mach-Zehnder
spectrometer \cite{subm}, respectively.

\section{Results and Discussion}

In Fig. \ref{refl} the frequency dependent reflectivity of
La$_{1-x}$Sr$_{x}$MnO$_{3}$ is shown for four doping levels up to
$x=0.2$. The sharp resonance features showing up in the region
between 100 and 1000~cm$^{-1}$ (3~-~30~THz) are due to phonon
excitations. They are most prominent in the insulating regions of
the phase diagram, i.e. for the low doping levels shown in Figs.
\ref{refl}(a) and (b) and for the higher temperatures at the
higher doping levels given in Figs. \ref{refl}(c) and (d). For a
detailed discussion of these phonon modes the reader is referred
to references
\cite{mayr1,ParaJMMM,mayrprom,kim,fedorov,abrashev,pao2,hart2,hart3,mayr2}.
For the insulating states, immediately following the last phonon
mode the reflectivity increases sharply with increasing frequency.
This can be ascribed to the onset of electronic excitations across
a band gap, which at higher frequencies leads to a peak in the
reflectivity close to $10^{4}$~cm$^{-1}$ ($\approx 1.3$~eV). Above
about $2\times 10^{4}$~cm$^{-1}$, the onset of a further
electronic excitation is observed for all doping levels shown in
Fig. \ref{refl}. It is clear from the electronic structure of the
doped manganites that below $10^{4}$~cm$^{-1}$ excitations between
and within the JT-split $e_{g}$ levels dominate. The
charge-transfer gap, corresponding to a transition between the
oxygen 2$p$- and the manganese $e_{g}$ states, is expected close
to $4 ~\textrm{eV} \approx 3\times10^{4} ~\textrm{cm}^{-1}$ and is
just covered at the high-frequency end of the reflectivity spectra
for all compounds shown in Fig. 2. For the metallic sample with
$x=0.2$ [Fig. \ref{refl}(d)] and when the metallic state is
approached with decreasing temperature at for $x=0.175$ [Fig.
\ref{refl}(c)], the gaplike excitations below $10^{4}$~cm$^{-1}$
and the phonon modes are superimposed by a strong additional
contribution, smoothly increasing and finally tending to saturate
with decreasing frequency. These are the characteristic features
of a Drude contribution due to free charge carriers.

\begin{vchfigure}[htb]
\includegraphics[width=0.7\textwidth,clip]{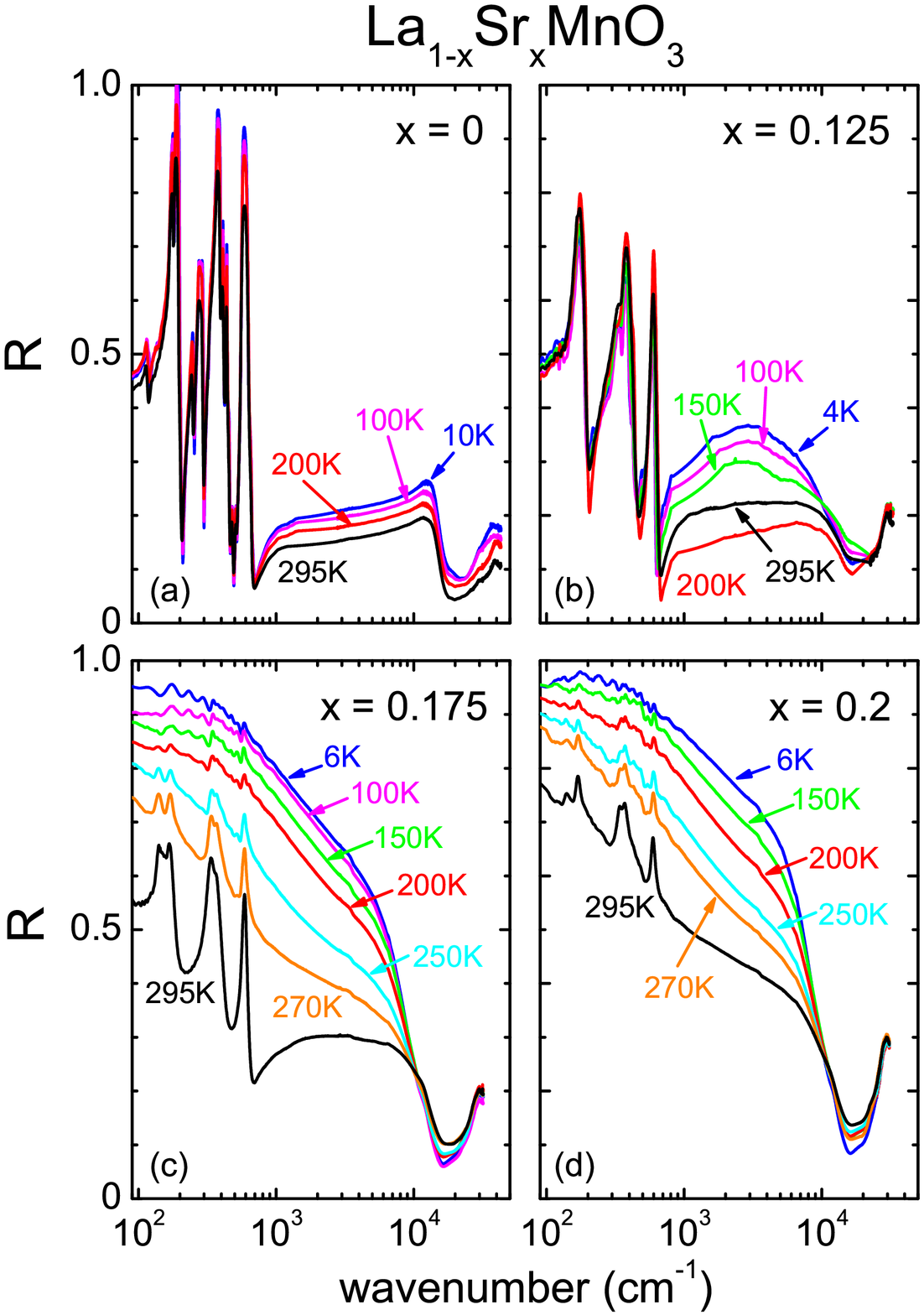}
\vchcaption{Frequency dependent infrared reflectivity of
La$_{1-x}$Sr$_{x}$MnO$_{3}$ for four different doping levels and
various temperatures.} \label{refl}
\end{vchfigure}

To gain deeper insight into the various processes contributing to
the infrared reflectivity and to relate the results to those
obtained at sub-GHz frequencies, it is helpful to calculate the
frequency dependent conductivity from the reflectivity using the
Kramers-Kronig transformation. Figure \ref{x0} shows the frequency
dependence of $\sigma ^{\prime}$ of pure LaMnO$_{3}$ at selected
temperatures, combining results at low frequencies \cite{seeger}
and in the infrared region \cite{mayrprom}. Only those parts of
the spectra are shown that are not affected by contact
contributions \cite{contact}, which were shown in \cite{seeger} to
dominate at very low frequencies and high temperatures. As already
noted in \cite{seeger}, for $\nu<1$~GHz, three contributions can
be identified: For $\nu\rightarrow0$, $\sigma^{\prime}$ approaches
the frequency-independent dc conductivity. Its temperature
dependence follows the prediction of the Variable-Range-Hopping
model \cite{seeger}. With increasing frequency, $\sigma^{\prime}$
increases smoothly, which can be parameterized by a sublinear
power law $\sigma^{\prime}\propto\nu^{s}$ with $s<1$
\cite{seeger}, a behavior termed "Universal Dielectric Response"
(UDR) due to its universal occurrence in a variety of materials
\cite{udr}. Such a behavior is commonly regarded as a hallmark
feature of hopping conduction of charge carriers subjected to
disorder-induced localization \cite{hopping}. Superimposed to this
power law, a shoulder shows up at low temperatures. As revealed by
the upper inset of Fig. \ref{x0}, it corresponds to
well-pronounced peaks in the dielectric loss
$\varepsilon^{\prime\prime}\propto \sigma^{\prime}/\nu$ shifting
through the frequency window with temperature. This is typical for
a relaxational process, which was ascribed to localized hopping of
polarons in \cite{seeger}. However, it should be noted that the
energy barrier of 86~meV, determined from the temperature
dependence of the relaxation time \cite{seeger}, obviously has no
relation to the energy scales of the polaronic excitations
observed in the infrared region, which are discussed below.

\begin{vchfigure}[htb]
\includegraphics[width=0.7\textwidth,clip]{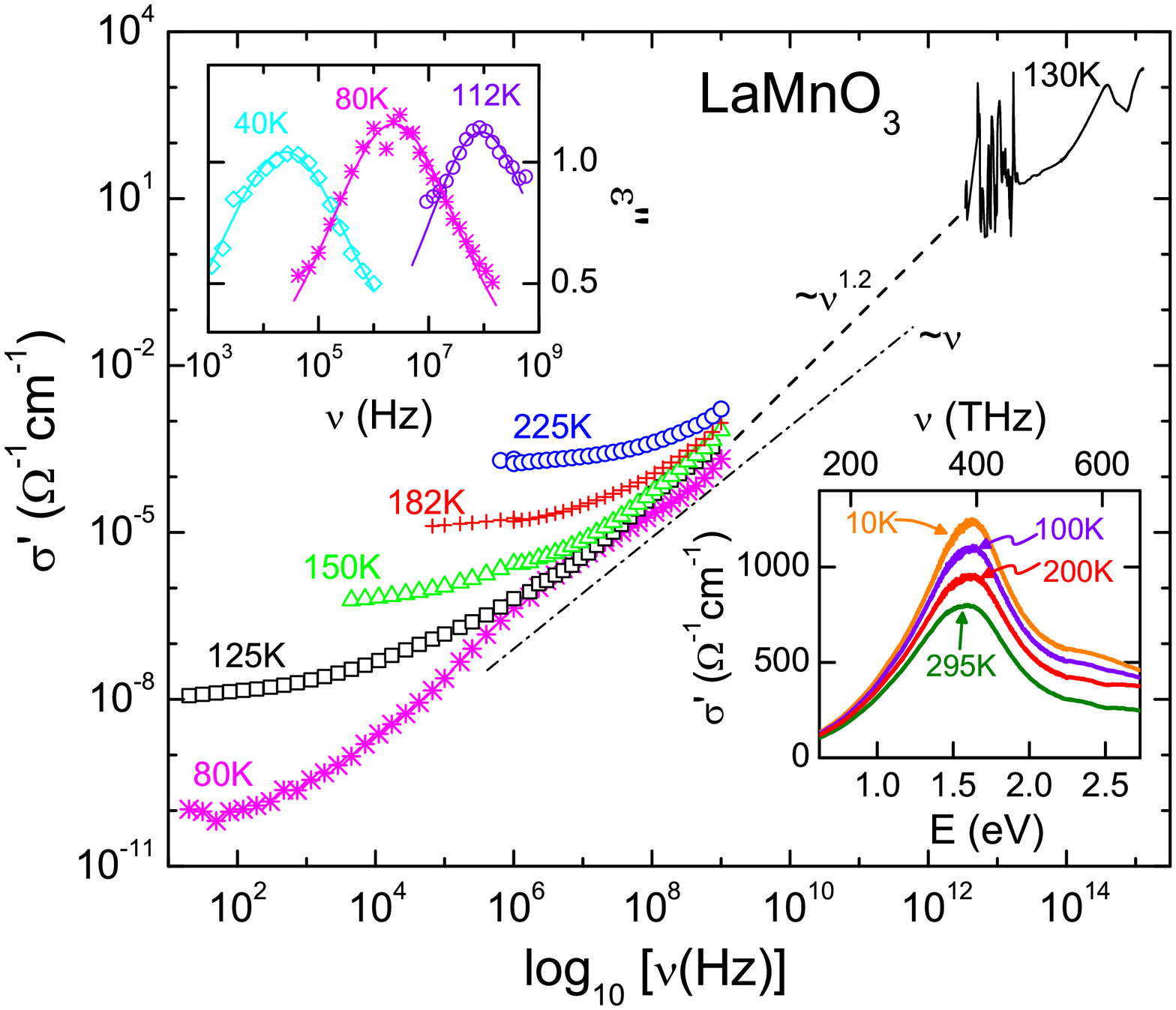}
\vchcaption{Frequency dependent conductivity of LaMnO$_{3}$ at
sub-GHz and infrared frequencies for various temperatures. The
upper inset shows the dielectric loss below GHz; the lines are
fits with the empirical Cole-Cole function \cite{seeger}. The
lower inset shows $\sigma ^{\prime}$ in the region of the
electronic excitation close to 1.5~eV for various temperatures.}
\label{x0}
\end{vchfigure}

It is not possible to extrapolate the sublinear power law found
below GHz towards the infrared results in the THz regime, even if
assuming $s=1$ as indicated by the dash-dotted line in Fig.
\ref{x0}. Clearly an additional contribution must be present in
the so-far uninvestigated intermediate region. We propose that
there is an additional superlinear power law (SLPL),
$\sigma^{\prime}\propto\nu^{n}$, with $n\approx1.2$ as indicated
by the dashed line. Such a SLPL was clearly detected, e.g., in the
CMR manganite Pr$_{0.65}$(Ca$_{0.8}$Sr$_{0.2}$)$_{0.35}$MnO$_{3}$
\cite{sichel} and evidenced to be a universal feature of
disordered matter \cite{univ}, e.g. supercooled liquids or doped
semiconductors. In the present case of a nominally undoped
material without any substitutional disorder, a slight
off-stoichiometry or impurities at the ppm level seem to be
sufficient to produce this typical response of disordered matter,
as it is also the case, e.g., for LaTiO$_{3}$ \cite{latio}.

In the infrared region, at frequencies beyond the phonon
resonances a peak shows up at about 380~THz, corresponding to
1.6~eV. As shown in the lower inset of Fig. \ref{x0}, its peak
frequency is nearly independent of temperature. This peak most
probably corresponds to an electronic transition between the
JT-split $e_{g}$ bands. The excitation from one $e_{g}$ orbital to
another at the same site is forbidden and should have negligible
oscillator strength. Hence, this peak structure must correspond to
a transition between Mn-ions on adjacent sites, which, due to
Hund's coupling, is favorable for parallel spin alignment. Thus,
at first glance at the magnetic phase transition an anomaly in the
intensity is expected, but experimentally only a moderate
temperature dependence is observed (lower inset of Fig. \ref{x0}).
This finding may be explained considering that in the paramagnetic
phase the spins are thermally disordered while in the type-A
antiferromagnetic state, order is established with ferromagnetic
spin alignment within and antiferromagnetic order between the
($a$,$b$)-planes. But the above interpretation is not corroborated
by polarized light-scattering experiments, which indicate a strong
charge-transfer character of this transition \cite{tobe}. Further
experiments will be necessary to unravel the nature of this
transition.

\begin{vchfigure}[htb]
\includegraphics[width=0.5\textwidth]{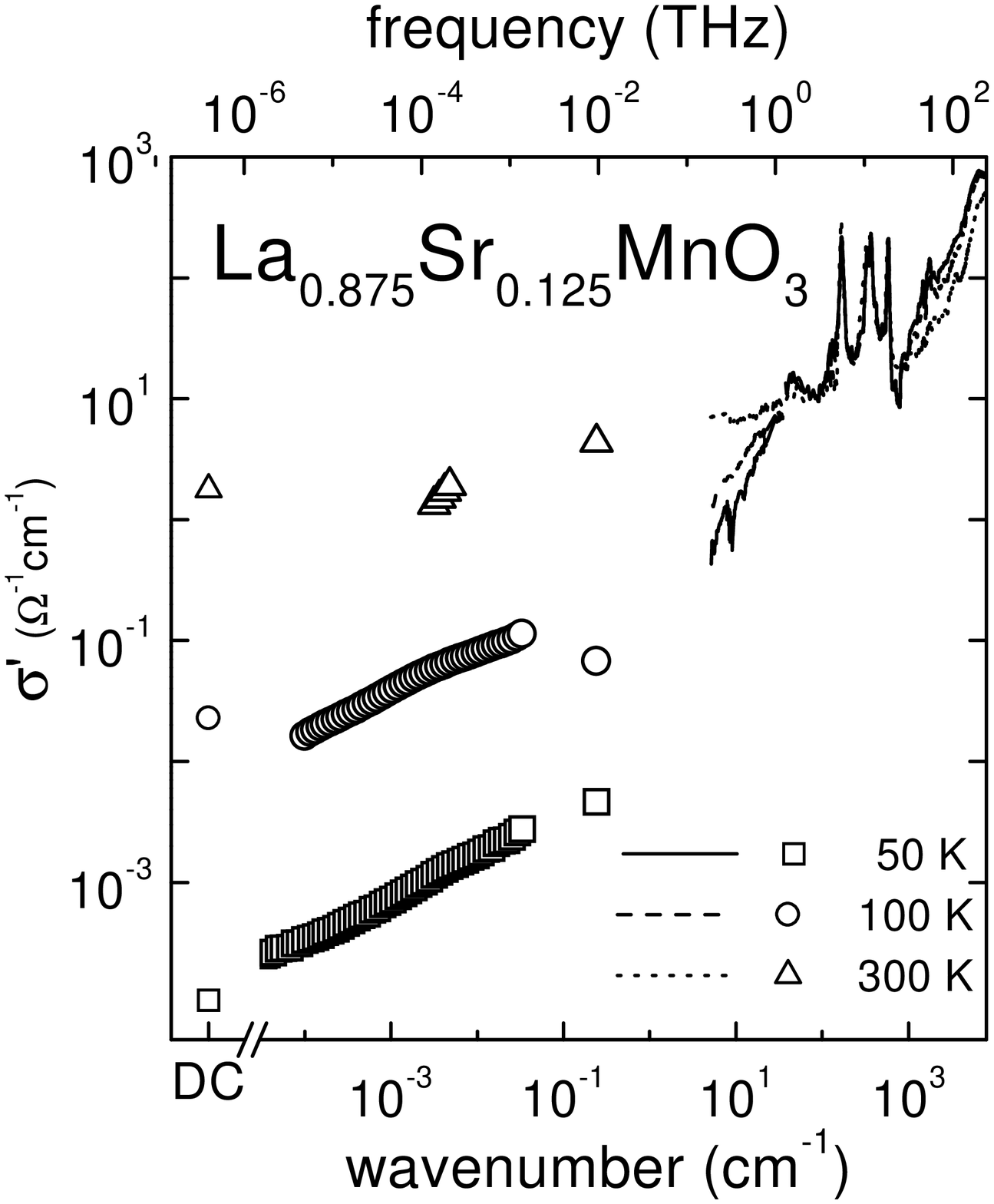}
\vchcaption{Frequency dependent conductivity of
La$_{0.875}$Sr$_{0.125}$MnO$_{3}$ at three temperatures. In
addition, the results of a four-point dc-measurement are
indicated. (reproduced from \cite{ParaJMMM})} \label{x125}
\end{vchfigure}

In Fig. \ref{x125}, reproduced from ref. \cite{ParaJMMM}, the
frequency-dependent conductivity of
La$_{0.875}$Sr$_{0.125}$MnO$_{3}$ is shown for three temperatures.
At room temperature, where the dc conductivity is rather high, the
low-frequency measurements provide information on the intrinsic
conductivity in a restricted frequency range only, because at low
frequencies the contact contributions and at high frequencies the
inductance of the sample dominates \cite{seeger}. Nevertheless,
from Fig. \ref{x125} it seems clear that in the sub-GHz range the
room-temperature conductivity is frequency independent. However,
interpolating from GHz to the lowest frequencies of the FIR
experiment, thereby taking into account the single point at
7.3~GHz from a microwave resonance measurement, it seems likely
that in this region a small increase of $\sigma^{\prime}(\nu)$ is
present. It can be suspected to be caused by hopping conduction of
localized charge carriers, the corresponding sublinear power law
emerging from the dc background with increasing frequency. The
presence of this contribution, pointing to a localization of the
charge carriers also in this compound, becomes even more obvious
at the lower temperatures, where a succession of a clearly
pronounced sublinear power law and a SLPL can be suspected. The
non-metallic character of this compound is further corroborated by
the complete absence of a Drude contribution in the infrared
region. Instead, above the phonon modes, the conductivity exhibits
a strong increase, which can be ascribed to superposed
contributions from polaron absorption and interband transitions,
as discussed in detail in \cite{mayr1,mayrprom}.

\begin{vchfigure}[htb]
\includegraphics[width=0.85\textwidth]{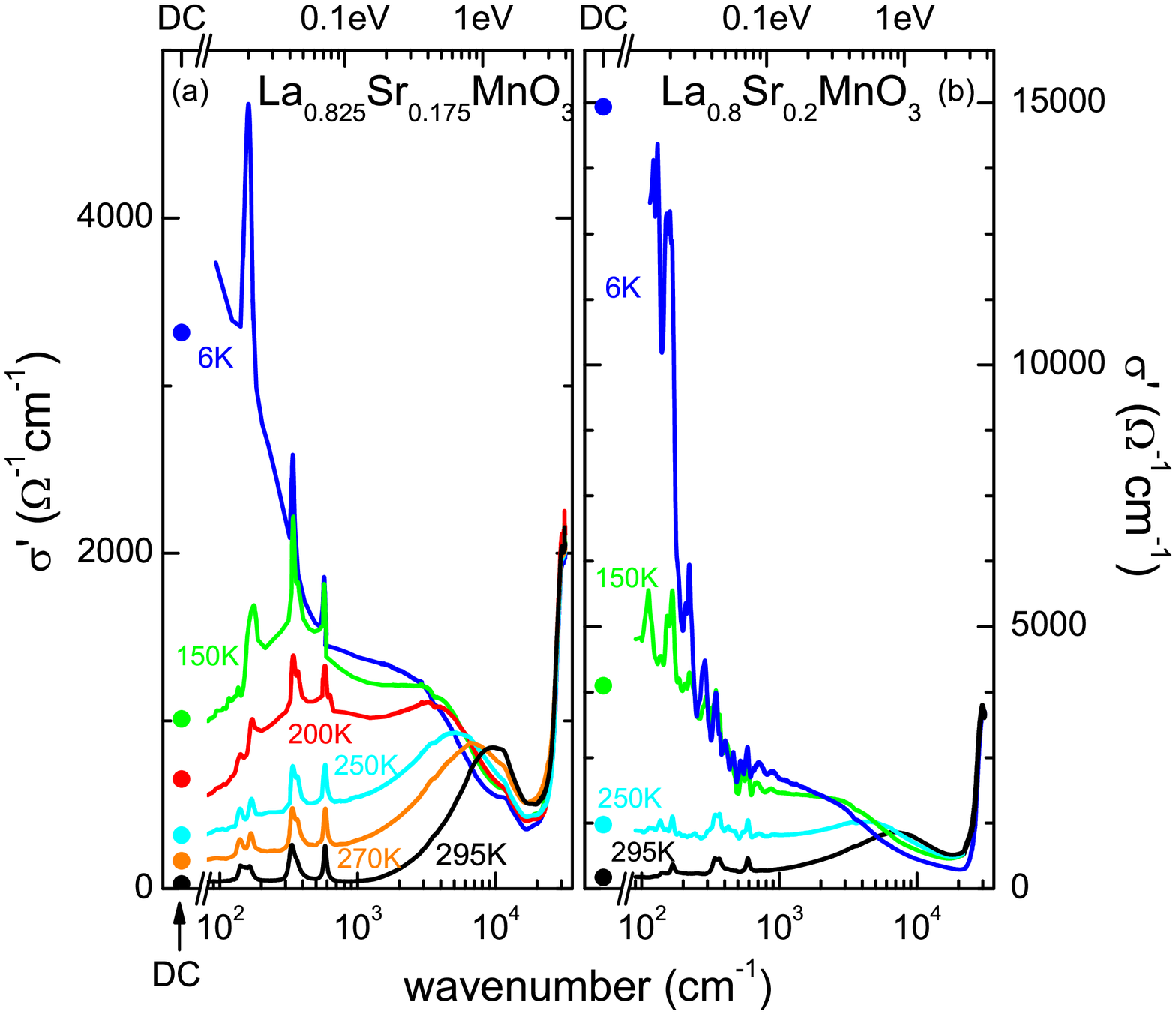}
\vchcaption{Frequency dependent conductivity of
La$_{1-x}$Sr$_{x}$MnO$_{3}$ for $x=0.175$ and 0.2 at various
temperatures. In addition, the results of a four-point
dc-measurement are indicated.} \label{xhigh}
\end{vchfigure}

For Sr contents $x=0.175$ and 0.2, having ferromagnetic metallic
ground states, the measured low-frequency conductivity is
dominated by inductance and skin effect contributions
\cite{seeger}. The intrinsic $\sigma^{\prime}(\nu)$ can be
expected to reveal a significant frequency dependence only at high
frequencies, in the infrared and optical region, as shown in Fig.
\ref{xhigh}. For $x=0.175$, at high temperatures, where the sample
is close to the MI transition (cf. Fig. \ref{PhaseDiag}),
$\sigma^{\prime}(\nu)$ shows a broad peak at about
$10^{4}$~cm$^{-1}$. With decreasing temperature, spectral weight
is shifted from this excitation towards lower frequencies, leading
to a peak or shoulder strongly shifting towards lower frequencies
when the sample becomes more metallic with decreasing temperature.
Finally, at 6~K $\sigma^{\prime}$ increases markedly towards low
frequencies, in good accord with the strongly enhanced dc
conductivity, thus exhibiting the clear signature of a Drude
contribution of free charge carriers. A detailed analysis of the
Drude response in La$_{0.825}$Sr$_{0.175}$MnO$_{3}$ was provided
in \cite{take}. It seems reasonable to assume that the excitation
at about 11000~cm$^{-1}$ or 1.4~eV has a similar origin as that at
1.6~eV for pure LaMnO$_{3}$. But now we have a fraction of
Mn$^{4+}$ sites with no $e_{g}$ electrons. Hence, in addition to
transitions between JT-split $e_{g}$ orbitals, optical transitions
between empty and occupied $e_{g}$ levels have to be taken into
account. We believe that the broad peak at room temperature at
1.4~eV is a mixture of both processes. With decreasing
temperatures a remainder of the JT-split derived peak still is
visible at 1.4~eV, indicating the significantly weaker JT energy
in the doped crystals compared to the 1.6~eV observed in the pure
compound (Fig. \ref{x0}). The second spectral feature branching
off from this excitation and being strongly temperature dependent
can be ascribed to a polaronic excitation. Similar behavior has
been reported in \cite{oki1} exemplifying the importance of strong
electron-phonon interactions in the manganites. The strong
softening of this polaronic excitation qualitatively resembles the
behaviour predicted by Millis \cite{millis1} for a scenario of
strong electron-phonon coupling. However, in the case of
half-filling, the polaronic excitation transforms into a Drude
type behavior at low temperatures. In La$_{1-x}$Sr$_{x}$MnO$_{3}$
($0 \leq x \leq 0.2$) a Drude peak evolves with doping
simultaneously with the polaronic excitation, with the Drude peak
being characterized by a low relaxation rate and a very low
optical weight. The rather symmetric shape of the polaron peak
with the absence of a clear cut-off frequency at low frequencies
indicates the dominance of small polarons \cite{hart1}. Finally,
at the high-frequency edge of the investigated spectral range, a
further strong increase of the conductivity shows up, indicating
the onset of the charge-transfer excitation as in pure
LaMnO$_{3}$. For $x=0.2$ [Fig. \ref{xhigh}(b)], the overall
behavior is qualitatively similar to that observed for $x=0.175$.
However, the excitation ascribed to the polaronic transition seems
to be shifted to lower energy, now located at about 0.8~eV at room
temperature and of significantly lower optical weight. At the
lowest temperatures a narrow Drude peak dominates the optical
conductivity and the polaronic excitation appears as a weak
shoulder close to 3000~cm$^{-1}$.

\begin{vchfigure}[htb]
\includegraphics[width=0.5\textwidth]{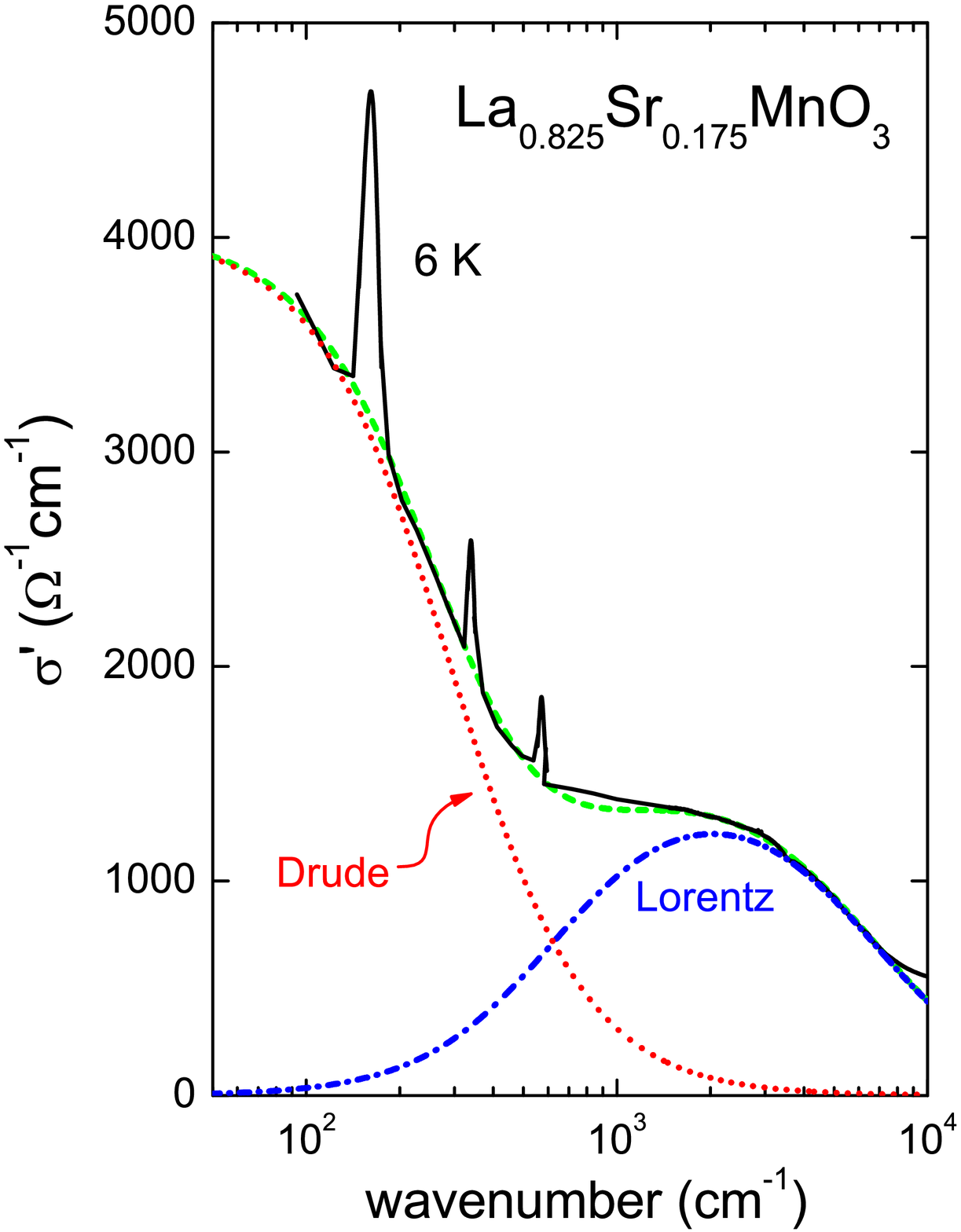}
\vchcaption{Frequency dependent conductivity of
La$_{0.825}$Sr$_{0.175}$MnO$_{3}$ at 6~K (solid line). The dashed
line is a fit of the experimental spectrum using the sum of a
Drude and a Lorentz term. These two contributions are indicated by
the dotted and dash-dotted lines, respectively.} \label{fits}
\end{vchfigure}

In Fig. \ref{fits}, a least-square fit of the conductivity of
La$_{0.825}$Sr$_{0.175}$MnO$_{3}$ measured at 6~K is shown (dashed
line). For the fit, the sum of a Drude term dominating at low
frequencies and a Lorentz term (peak frequency $\nu
_{0}=2040~\textrm{cm}^{-1}$, damping constant $\gamma =
7180~\textrm{cm}^{-1}$) taking into account the polaronic
excitation was used. The three phonon modes were neglected. A good
description of the experimental data could be achieved in this
way, revealing a Drude relaxation time of
$\tau=1.8\times10^{-14}$s. Using a Fermi velocity for the
orthorhombic structure of $2.2\times10^{7}$~cm/s \cite{singh}, we
obtain a mean free path of 4~nm, which characterizes this compound
as a bad metal. As discussed earlier, the symmetric form of the
polaron absorption points towards the existence of small polarons
\cite{hart1}. The small ratio of the optical weight of the Drude
conductivity to that of the polaron absorption (Fig. \ref{fits})
indicates that only a small fraction of free charge carriers
contributes to the Drude-like conductivity, while the majority
still is localized by strong electron-phonon coupling effects.

\section{Summary and Conclusions}

The ac conductivity of the CMR manganite
La$_{1-x}$Sr$_{x}$MnO$_{3}$ was reported for four doping levels
$x$. The combination of low-frequency and optical measurements
revealed the importance of hopping conduction of localized charge
carriers for the non-metallic parts of the phase diagram of this
system. In the infrared region beyond the phonon modes, evidence
for two electronic modes were obtained, one of them showing the
clear signature of a polaronic excitation. Deep in the metallic
state a strong Drude contribution is observed.

\begin{vchfigure}[htb]
\includegraphics[width=0.6\textwidth]{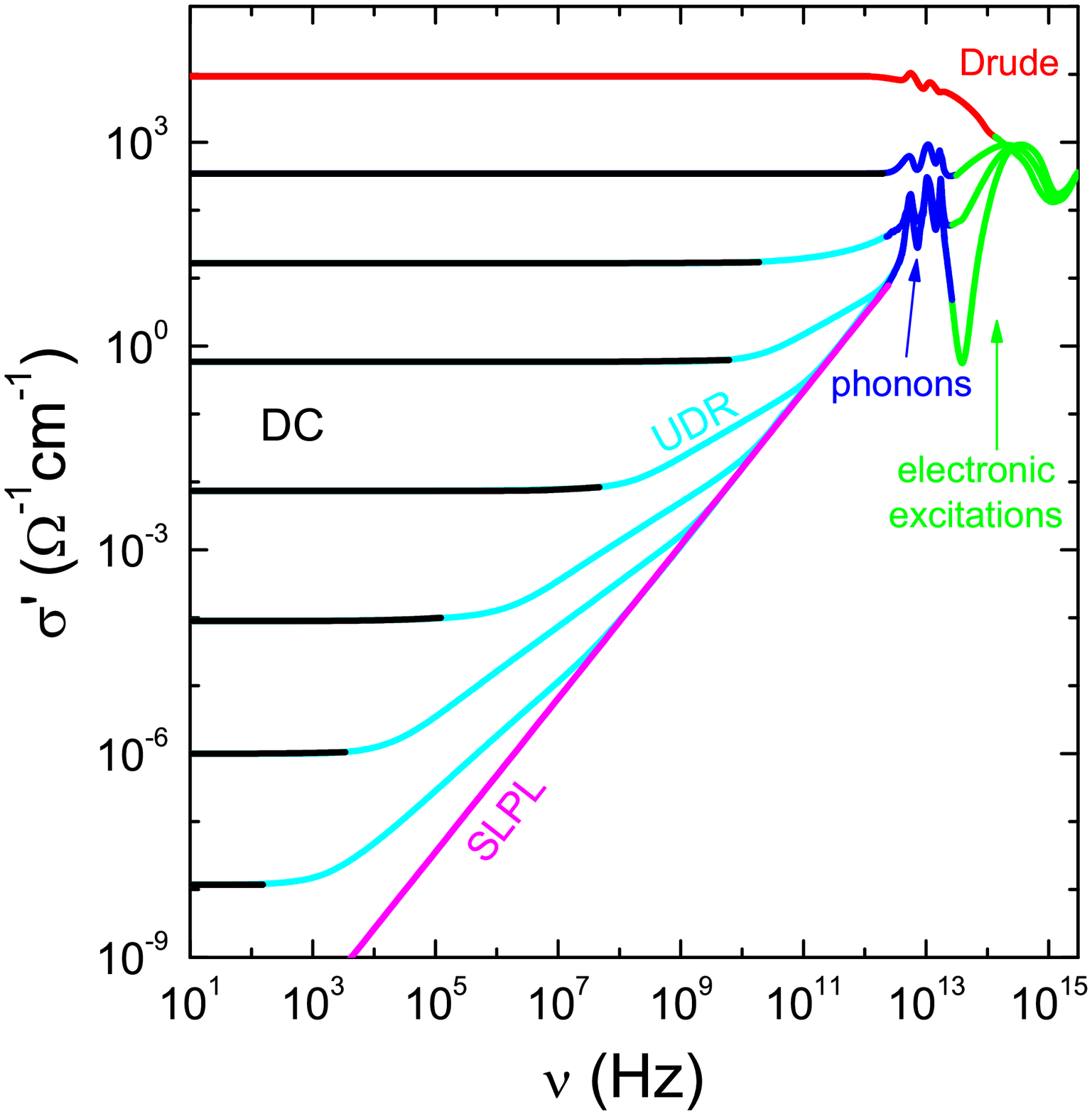}
\vchcaption{Schematic plot of the conductivity of a system
approaching a MI transition from the insulating side as observed,
e.g., in transition metal oxides. The figure covers a broad range,
starting at frequencies of typical dielectric experiments up to
those covered by optical spectrometers. The lowest curves
correspond to the most insulating, the highest to the most
metallic states. The values of the tick labels are intended to
give a rough estimate only.} \label{scheme}
\end{vchfigure}

Taking together the results of the present work and those obtained
on a variety of further transition metal oxides and other systems
being close to a MI transition
\cite{seeger,sichel,univ,latio,broadband}, allows for the
development of a general scenario for the broadband ac
conductivity when approaching the MI transition from the
insulating side. As schematically indicated in Fig. \ref{scheme},
deep in the insulating phase most materials show a succession of a
dc plateau, a sublinear, and a superlinear power law \cite{univ}.
Such a behavior seems to be closely connected to hopping
conduction of charge carriers that are localized due to disorder.
One should note, however, that this notion is mainly based on
theories that were developed for amorphous or heavily doped
conventional semiconductors \cite{hopping,hoptheo} and whose
predictions can be approximated by a sublinear power law. It is
not clear, if these theories also should be applicable to
electronically correlated materials as, e.g., the CMR manganites.
Also so far there is no well-founded explanation for the SLPL,
which seems to occur concomitantly with the sublinear law.
Whatsoever, it seems clear now that in the insulating and
semiconducting states these power laws are also active in the only
rarely investigated transition region between the classical
dielectric frequencies ($\sim$ below GHz) and those of typical
infrared experiments. Often they even seem to represent the
background for the phonon modes, however, vanishing abruptly
beyond the highest phonon resonance. When the system further
approaches the metallic state, the dc conductivity, growing more
rapidly than the UDR and the nearly constant SLPL, dominates over
an increasingly broader frequency range, which finally extends
well up to the first phonon modes (second curve from above in Fig.
\ref{scheme}). This typically occurs for a dc conductivity of
about 100~-~1000~$\Omega^{-1}$cm$^{-1}$ and marks the transition
into the metallic state. Up to this point, the infrared response
is dominated by strong phonon modes, followed by electronic
excitations due to interband and polaronic contributions. Within
the metallic phase, the phonons and low-lying electronic modes are
successively superimposed by the typical free carrier Drude
behavior as indicated by the uppermost curve in Fig. \ref{scheme}.
It is interesting that the value of $\sigma^{\prime}$ marking the
transition from insulating to a Drude-like metallic behavior seems
to be rather universal and a connection of this finding to the
concept of the "minimum metallic conductivity" proposed by Mott
\cite{mott} may be suspected.

\begin{acknowledgement}
We thank A.A. Mukhin and A.M. Balbashov for providing the samples
and A. Pimenov, K. Pucher, and A. Seeger for performing part of
the measurements. Stimulating discussion with Th. Kopp and Ch.
Hartinger are gratefully acknowledged. This work was supported by
the Deutsche Forschungsgemeinschaft via the
Sonderforschungsbereich 484 and by the BMBF via VDI/EKM.
\end{acknowledgement}

\end{document}